\documentstyle [aps,twocolumn]{revtex}
\catcode`\@=11
\newcommand{\be}{\begin{equation}}
\newcommand{\ee}{\end{equation}}
\newcommand{\bea}{\begin{eqnarray}}
\newcommand{\eea}{\end{eqnarray}}

\newcommand{\up}{\uparrow}
\newcommand{\down}{\downarrow}

\newcommand{\ri}{\mbox{i}}
\newcommand{\re}{\mbox{e}}

\begin {document}
\title{Composite Pairings in Chirally Stabilized Critical Fluids} 
\vspace{2cm}

\author{P. Azaria$^1$ and   P. Lecheminant$^2$}

\vspace{0.5cm}

\address{$^1$ Laboratoire de Physique Th\'eorique des Liquides,
Universit\'e Pierre et Marie Curie, 4 Place Jussieu, 75252 Paris,
France\\
$^2$ Laboratoire de Physique Th\'eorique et Mod\'elisation,
Universit\'e de Cergy-Pontoise, 5 Mail Gay-Lussac, 
Neuville sur Oise, 95301
Cergy-Pontoise Cedex, France}

\vspace{3cm}

\address{\rm (Received: )}
\address{\mbox{ }}
\address{\parbox{14cm}{\rm \mbox{ }\mbox{ }
We study a one-dimensional electron gas in a special
antiferromagnetic environment  
made by two spin-1/2 Heisenberg chains
and the one-dimensional 
two-channel Kondo-Heisenberg lattice away from half-filling.
These models flow to an intermediate fixed point
which belongs to the universality class of chirally stabilized liquids.
Using a Toulouse point approach,
the universal properties of the models
are determined as well as the identification of
the leading instabilities.
It is shown that these models exhibit 
a non-Fermi liquid behavior with strong enhanced composite pairing
correlations.
}}
\address{\mbox{ }}
\address{\parbox{14cm}{\rm PACS No: 75.10.Jm, 75.40.Gb}}
\maketitle

\makeatletter
\global\@specialpagefalse
\makeatother

The concept of odd-frequency superconducting 
ordering was conceived by Berezinskii\cite{berezinski}
who considered spin-triplet odd-frequency pairing 
as an alternate of the conventional theory
$^{3}$He pairing. The possibility of such unusual pairing 
was revived nearly one decade ago by Balatsky and 
Abrahams\cite{balatsky} in the context of 
singlet superconductors 
and by Emery and Kivelson\cite{emery} in the two-channel 
Kondo problem with the emphasis laid on the composite nature 
of the order parameter for odd-frequency pairing.
This idea of composite pairing was further developped 
by Coleman et al.\cite{coleman}
who proposed that heavy-fermion superconductors might involve 
odd-frequency triplet pairing.

One of the most important difficulty in finding some specific lattice 
models for realization of odd-frequency pairing
stems from the fact that it requires a controlled
analysis in the strong coupling regime. 
Such an approach is possible in some extreme limits as in one dimension
where non-perturbative techniques are available.
Strong odd-frequency singlet pair correlations were, in particular, 
identified within the bosonization 
approach\cite{book} 
of the one-dimensional single channel Kondo lattice
in an anisotropic limit\cite{zachar,georges}.
Another candidate is the one dimensional single 
channel Kondo-Heisenberg
lattice (KHL) which consists of a one dimensional electron gas (1DEG) 
interacting
with an antiferromagnetic Heisenberg spin-1/2 chain 
by a Kondo coupling. 
Away from half filling, this model
has a spin gap\cite{white} and is thus expected to 
exhibit enhanced pairing correlations.  
In particular, it has been shown recently\cite{zachar1}
that the dominant instabilities are
of a composite nature in a certain regime of the parameters. 
In this letter, we shall study two different 
generalization of the one dimensional KHL 
with two channels which have no spin gap 
{\sl but} exhibit dominant unconventional pairing instabilities.
A first generalization is to consider 
a 1DEG coupled by a Kondo 
coupling to two non-interacting
antiferromagnetic Heisenberg spin-1/2 chains. 
This model belongs to the more general class of Luttinger 
liquids in active environments introduced in the context of the striped 
physics\cite{emery1,neto,granath}.
A second generalization consists of 
conduction electrons with a two-fold orbital degeneracy interacting
with a periodic array of localized spins i.e. the
two-channel KHL.
It has been shown in Ref.\cite{azarial} that these two models 
away from half filling exhibit a critical phase governed by
a fixed point which belongs to
the class of chirally stabilized fluids\cite{andrei}.
In this work, we shall compute  correlation functions of physical
observables by means of a Toulouse point solution and characterize 
the dominant instabilities of both models which turn out to be of a 
composite nature. Very recently, the one dimensional N-channel KHL
 has been investigated
by Andrei and Orignac\cite{orignac} using a conformal field theory (CFT) 
approach. These authors found that the leading instability is of
a composite pairing type when $N \le 4$\cite{orignac}. In this respect,
the exact solution at the Toulouse point provides an independent and 
physically transparent approach for the particular 2-channel KHL.

{\sl The models.} 
The first model (referred in the following as 
a 1DEG in a special active environment) consists of a 1DEG 
away from 
half filling coupled symmetrically
by a Kondo coupling ($J_K >0$) with two non-interacting 
antiferromagnetic ($J_H >0$) spin-1/2 Heisenberg chains:
\bea 
{\cal H}_1 &=& -t \sum_{i} \left(c^{\dagger}_{i\sigma}
c_{i+1\sigma} + H.c.\right) 
+J_K \sum_{a=1}^{2}\sum_{i}{\bf S}_{c i} \cdot {\bf S}_{a i} 
\nonumber \\
&+& J_H \sum_{a=1}^{2} \sum_{i}{\bf S}_{a i} \cdot {\bf S}_{a i+1}.
\label{model1}
\eea
Here, $c_{i\sigma}$ is the conduction electron operator at site $i$
with spin $\sigma= \up, \down$,  
${\vec S}_{c i} =
c^{\dagger}_{i\alpha} {\vec {\sigma}}_{\alpha\beta}
c_{i\beta}/2$ (${\vec \sigma}$ being the Pauli matrices)
stands for the electron spin operator at site $i$
whereas
the localized spin operator
at site $i$ on the chain of index $a=1,2$ is denoted by ${\bf S}_{a i}$.
In the limit $J_K\ll t, J_H$, 
the low energy behavior of the model can 
be determined using the continuum limit of the electron operator
in terms of right and left moving fermion fields:
$c_{i\sigma} \rightarrow R_{\sigma} {\re}^{ik_F x}
+ L_{\sigma} {\re}^{-ik_F x}, x=i a_0$ ($k_F$ being the Fermi
momentum and $a_0$ the lattice spacing).
The continuum description of the conduction
spin density operator ${\bf S}_c$ can
then be expressed in terms of a spin current ${\bf J}_{0R,L}$
belonging to the SU(2)$_1$ Kac-Moody (KM) algebra\cite{book}
and a bosonic field $\Phi_c = \Phi_{cL}
+ \Phi_{cR}$ that accounts for the charge degrees
of freedom:
${\bf S}_c \rightarrow {\bf J}_{0R} + {\bf J}_{0L} +
\cos(2 k_F x + \sqrt{2\pi} \Phi_c) {\bf n}_0$,
${\bf n}_0$ being the staggered magnetization.                 
In the same way,
the spin densities of the surface chains are represented 
as (see Refs. \cite{affleck,book}): 
${\bf S}_a (x) = {\bf J}_a (x) + (-1)^{x/a_0} {\bf n}_a(x)$ 
where ${\bf J}_a = {\bf J}_{aR} + {\bf J}_{aL}$
and ${\bf n}_a$ are respectively the uniform and 
staggered parts of the magnetization.
The chiral spin currents ${\bf J}_{aR,L}$ belong also to the 
SU(2)$_1$ KM algebra.
With this low energy description, the Hamiltonian density
of the lattice model (\ref{model1}) for 
incommensurate filling reads as follows in the
continuum limit:
\bea
{\cal H}_1 &\simeq&  
\frac{v_F}{2}\left(\left(\partial_x \Phi_c\right)^2
+ \left(\partial_x \Theta_c\right)^2\right) 
+\frac{2\pi v_F}{3}\left({\bf J}_{0R}^2
+ {\bf J}_{0L}^2\right) 
\nonumber \\
&+& \frac{2\pi v_H}{3}\sum_{a=1}^{2}\left({\bf J}_{aR}^2
+ {\bf J}_{aL}^2\right)
+ g \left({\bf J}_{0R} \cdot {\bf I}_{L} 
+ {\bf J}_{0L} \cdot {\bf I}_{R}\right)
\label{contmodel1}
\eea
where we have neglected all oscillatory,
marginal irrelevant contributions as well as current-current interactions
of the same chirality. As shown in Ref.\cite{azarial}, at the strong
coupling fixed point, these latter interactions only lead to 
renormalization of velocities and logarithmic corrections. 
In Eq. (\ref{contmodel1}), 
$\Theta_c = \Phi_{cL} - \Phi_{cR}$ is the dual charge field,
$v_F$ the Fermi velocity, 
$v_H$ the spin velocity of the magnetic
environment, and $g\simeq J_Ka_0 >0$ is the interacting coupling constant;
${\bf I}_{R,L} = {\bf J}_{1R,L} + {\bf J}_{2R,L}$ 
is a SU(2)$_2$ KM current 
being  the sum of two SU(2)$_1$ KM currents.

The second model considered in this letter is the 
one-dimensional two-channel KHL:
\bea
{\cal H}_2 &=& -t \sum_{i} \sum_{a=1}^{2}\left(c^{\dagger}_{i a\sigma}
c_{i+1 a\sigma} + H.c.\right) 
+ J_H \sum_{i}{\bf S}_{0 i} \cdot {\bf S}_{0 i+1}
\nonumber \\
&+& J_K \sum_{a=1}^{2}\sum_{i}{\bf S}_{ac i} \cdot {\bf S}_{0 i}
\label{model2}                                                         
\eea
where now the electron operator $c_{ia\sigma}$ has a channel index $a=1,2$
and interacts by a 
Kondo coupling ($J_K >0$) with a periodic array of 
localized spins ${\bf S}_{0 i}$. 
Away from half filling,
the continuum limit of the Hamiltonian (\ref{model2})  
proceeds in the same way as in the first model and
with the same degree of approximations than in Eq. (\ref{contmodel1}), 
one has:
\bea
{\cal H}_2 &\simeq&
\frac{v_F}{2}\sum_{a=1}^{2}\left(\left(\partial_x \Phi_{ac}\right)^2
+ \left(\partial_x \Theta_{ac}\right)^2\right)
+\frac{2\pi v_H}{3}\left({\bf J}_{0R}^2
+ {\bf J}_{0L}^2\right) \nonumber\\
&+& \frac{2\pi v_F}{3}\sum_{a=1}^{2}\left({\bf J}_{aR}^2
+ {\bf J}_{aL}^2\right)
+ g \left({\bf J}_{0R} \cdot {\bf I}_{L}
+ {\bf J}_{0L} \cdot {\bf I}_{R}\right)
\label{contmodel2}
\eea                                              
where $\Phi_{ac}$ is a bosonic field ($\Theta_{ac}$
being the dual field) associated with 
charge fluctuations in the ath channel.
The chiral SU(2)$_1$ currents ${\bf J}_{aR,L}$ 
correspond to the chiral
uniform part of the electron spin density whereas
${\bf J}_{0R,L}$ are the chiral SU(2)$_1$ currents of
the local moments.

{\sl Toulouse point solution.}
The next step of the approach is to use the representation of
two  SU(2)$_1$ currents in terms of four Majorana fermions 
$\xi^{0}$ and ${\vec \xi}$\cite{book,allen}:
${\bf I}_{R,L}
= -\ri \; {\vec \xi}_{R,L} \wedge {\vec \xi}_{R,L}/2,
{\bf J}_{1R,L} - {\bf J}_{2R,L} = \ri
{\vec \xi}_{R,L} \xi_{R,L}^{0}$.
The Hamiltonians (\ref{contmodel1},\ref{contmodel2}) 
share then the same structure:
${\cal H}_{1,2} = {\cal H}^{0}_c + {\cal H}^{0}(\xi^0) + \bar{\cal H}$ 
where ${\cal H}^{0}_c$ is the free Hamiltonian 
of the charge bosonic fields and 
${\cal H}^{0}(\xi^0)$ is the free Hamiltonian 
of the massless Majorana fermion $\xi^0$.
All nontrivial physics is incorporated in $\bar{\cal H}$ 
which separates into two commuting and chirally asymmetric
parts: $\bar{\cal H} = \bar{\cal H}_1 + \bar{\cal H}_2,
\; [\bar{\cal H}_1, \bar{\cal H}_2] =0$ with
\be
\bar{\cal H}_1 =  \frac{\pi v_1}{2} \;{\bf I}_{R}^2 
+ \frac{2\pi v_0}{3} \; {\bf J}_{0L}^2 + g {\bf J}_{0L} \cdot {\bf I}_{R} 
\label{ham1}
\ee
with $v_1 = v_H, v_0 = v_F$ for the 1DEG in a special
active environment and the reverse
for the two-channel KHL;
$\bar{\cal H}_2$  is obtained from $\bar{\cal H}_1$ by
inverting chiralities of all the currents.
For an antiferromagnetic Kondo coupling ($g>0$), the interacting part of
$\bar{\cal H}$ is marginally relevant. 
Naively one will thus expect that the model will enter a massive
strong coupling region.
However, 
due to the chiral asymmetry of $\bar{\cal H}$,  
the effective interaction
flows towards a conformally invariant intermediate 
fixed point with a smaller central charge than 
in the ultraviolet. Stated differently,
the interaction produces a mass gap in some sectors 
but there are still some degrees of freedom that remain
critical in the infrared (IR). 
The nature of the fixed point of chirally asymmetric 
current-current models with the structure (\ref{ham1})
have been determined by Andrei et al.\cite{andrei}
by a combination of CFT and Bethe ansatz techniques.
This fixed point belongs to the class of chirally 
stabilized fluids introduced in Ref.\cite{andrei} describing
the IR behavior of one-dimensional interacting chiral fermions.
A simpler way to identify the critical degrees of freedom 
of model (\ref{ham1}) in the IR limit can be done by means
of a Toulouse point approach
as in the two-channel Kondo model\cite{emery}.
By allowing anisotropic interaction ($g \rightarrow
g_{\parallel}, g_{\perp}$), the  U(1) version of 
the Hamiltonian $\bar{\cal H}$ can be mapped onto 
free Majorana fermions for a special value of the interaction
$g_{\parallel}^*$. Although the position of the solvable point
is non-universal, the Toulouse limit solution captures 
the physical and universal properties of the models
including the SU(2) case. 
The details of the approach can be found in Refs.\cite{azaria1,azarial}
and we shall now briefly review it to fix the notations.
The starting point of the Toulouse solution is the Abelian bosonization   
of all SU(2)$_1$ currents ${\bf J}_{a}, a=0,1,2$ in terms of 
massless bosonic fields ${\varphi}_a$\cite{book}: 
$J_{aR,L}^z = (1/\sqrt{2\pi}) \partial_x \varphi_{aR,L}$, $ \;
J_{aR,L}^{+} = (1/2\pi a_0)\re^{\mp \ri \sqrt{8\pi}
\varphi_{aR,L}}$.
Introducing the symmetric combination of the fields:
$\varphi_{+} = (\varphi_1 + \varphi_2)/\sqrt{2}$,
the SU(2)$_2$ current ${\bf I}$ writes: 
$I^z _{R,L} = (1/\sqrt{\pi})\partial_x \varphi_{+R,L}$,   
$I_{R,L}^{+} = (\ri/\sqrt{\pi a_0})\xi_{R,L}^3 \kappa
\re^{\mp \ri \sqrt{4\pi} \varphi_{+R,L}}$. A
fermionic zero-mode operator $\kappa$ has been introduced
to ensure the proper anticommutation relations.
The following canonical transformation is then performed:
\bea
\varphi_0 &=& \sqrt{2} \; \bar \Phi_2 - \bar
\Phi_1, \;
\varphi_+ = \sqrt{2} \; \bar \Phi_1  - \bar \Phi_2
 \nonumber \\
\vartheta_0 &=& \sqrt{2} \; \bar \Theta_2 + \bar
\Theta_1, \;
\vartheta_+ = \sqrt{2} \; \bar \Theta_1  + \bar
\Theta_2
\label{cano}
\eea
where $\vartheta_0$ and $\vartheta_+$ (respectively $\bar \Theta_1$ and $\bar
\Theta_2$)
are the dual fields
associated with $\varphi_0$
and $\varphi_+$ (respectively $\bar \Phi_1$ and $\bar \Phi_2$).
For a special positive value (Toulouse point)
of $g_{\parallel},~(g^*
_{\parallel} =
4 \pi (v_0 + v_1)/3)$,
the arguments of the interacting terms
become those of free fermions so that they can be refermionized 
further
with the introduction of 
a pair of Majorana fields ($\eta,\zeta$) using
the correspondence: $\eta_{R,L} + \ri \zeta_{R,L}
= 
           (\kappa / \sqrt{\pi a_0}) \;
\re^{\pm \ri \sqrt{4\pi} \bar \Phi_{2R,L}}$.
One finally ends with:
\bea
\bar{\cal H} &=& \frac{u_1}{2} \left[  (\partial_x {\bar \Phi}_1)^2 +
 (\partial_x  {\bar \Theta}_1)^2 \right] - \frac{\ri u_2}{2}
 \left[ \zeta_R\partial_x \zeta_R - \zeta_L\partial_x
\zeta_L\right]\nonumber\\
&-& \frac{\ri v_1}{2}\left[ \xi_R^{3}\partial_x \xi_R^{3}
-\xi_L^{3}\partial_x \xi_L^{3}\right]
- \frac{\ri u_2}{2}
 \left[ \eta_R\partial_x \eta_R - \eta_L\partial_x
\eta_L\right]\nonumber\\
 &+&\ri m \left[ \xi_R^{3}\eta_L - \eta_R\xi_L^{3}\right]
\label{hfin}
\eea
where $m = g_{\perp}/2 \pi a_0$,
$
u_1 = (2 v_1 - v_0)/3$, and $u_2 = (2 v_0 - v_1)/3$.
The Toulouse solution is stable provided that all velocities
are positive i.e. $1/2 \le v_0/v_1 \le 2$.
The first two terms in Eq. (\ref{hfin}) describe 
decoupled free massless bosonic and Majorana fields,
$\bar \Phi_1$ and $\zeta$, contributing to criticality
with the central charge:
$
c =3/2.
$
The remaining part of the Hamiltonian (\ref{hfin}) has a spectral gap $m$
and describes hybridization of the Majorana
$\xi^3$ and $\eta$ fermions with different chiralities. 
Adding the contribution of the excitations that do not
participate in the interaction,  the total central charge in the IR
of the 1DEG in a special active 
environment  (respectively the two-channel KHL) is $c=3$ 
(respectively $c=4$).
Apart from the charge degrees of freedom,
the elementary excitations of the models in the IR are of 
different nature:
The magnetic excitations are 
spinons defined as 
$\sqrt{\pi/2}$ kinks of the bosonic field $\bar \Phi_1$ 
describing an effective S=1/2 Heisenberg chain  
and nonmagnetic singlet excitations associated with 
the two massless Majorana fermions $\xi^0, \zeta$ (critical Ising
degrees of freedom) referred as pseudocharge degrees of 
freedom in Refs.\cite{azaria1,azarial}.
These $Z_2$ excitations play a crucial
role in the nontrivial IR physical properties of the models
as we shall see now. 

{\sl Physical properties of the 1DEG in a special active environment.} 
The Green's function for the right-left moving fermions
can be computed using the bosonic representation:
$(R,L)_{\sigma} = (\kappa_{\sigma}/{\sqrt{2\pi a_0}})
\; \re^{\ri\tau_{\sigma}\pi/4} 
\re^{\pm \ri \sqrt{4\pi} \varphi_{\sigma R,L}}$
where $\tau_{\up,\down} = \pm 1$, 
$\varphi_{\sigma R,L} = (\Phi_{cR,L} 
+ \tau_{\sigma} \varphi_{0R,L})/\sqrt{2}$, and 
$\kappa_{\sigma}$ are Klein-factors to
insure the correct anticommutation relations between
fermion fields of different spin index\cite{comment}.
Using the Toulouse basis (\ref{cano}), one then obtains
at the IR fixed point the estimate:
\bea 
\langle R_{\sigma} \left(x,\tau\right) 
R_{\sigma^{'}}^{\dagger} \left(0,0\right) \rangle \sim \nonumber \\
\frac{\delta_{\sigma, \sigma^{'}}}{\left(v_F \tau - \ri x\right)^{1/2} 
\left(u_1 \tau + \ri x\right)^{1/2} \left(u_2\tau - \ri x\right)}
\label{greenmod1}
\eea
which means that the system displays non-Fermi liquid properties; 
the left-moving Green's function is obtained from 
Eq. (\ref{greenmod1}) by the substitution:
$\ri \rightarrow -\ri$. The most interesting physical quantities
are the correlation functions of the various possible order
parameters. The spin-spin correlation functions have been computed 
in Ref.\cite{azarial} and the slowest ones are the staggered
parts that decay as $x^{-3/2}$.
Using the 
previous Abelian bosonization of the chiral fermions
and the Toulouse basis (\ref{cano}), we find in the electronic sector
the following representation 
for the charge density
wave (CDW), spin density wave (SDW), singlet (SS) and triplet (ST)
superconducting order parameters
in terms of the different critical fields at the IR fixed point:
\bea 
{\cal O}_{CDW} = L_{\sigma}^{\dagger} R_{\sigma}
\simeq \re^{\ri \sqrt{2\pi} \Phi_{c}} 
\cos\left(\sqrt{2\pi}\bar \Phi_1\right) \zeta_L \zeta_R
\nonumber \\
{\vec {\cal O}}_{SDW} = 
 L_{\alpha}^{\dagger} {\vec \sigma}_{\alpha \beta}R_{\beta}
\simeq
\ri \re^{\ri \sqrt{2\pi} \Phi_{c}} \zeta_R \zeta_L \nonumber \\
\left[\cos\left(\sqrt{2\pi}\bar \Theta_1\right),
\sin\left(\sqrt{2\pi}\bar \Theta_1\right), 
\sin\left(\sqrt{2\pi}\bar \Phi_1\right)\right] 
\nonumber \\ 
{\cal O}_{SS} = -\ri L_{\alpha}
\left(\sigma^y\right)_{\alpha \beta}
R_{\beta} \simeq 
\re^{-\ri \sqrt{2\pi} \Theta_{c}} 
\cos\left(\sqrt{2\pi}\bar \Phi_1\right) \zeta_R \zeta_L  
\nonumber \\ 
{\vec {\cal O}}_{TS} = -\ri L_{\alpha} 
\left({\vec \sigma} \sigma^y\right)_{\alpha \beta}
R_{\beta} \simeq 
\ri \re^{-\ri \sqrt{2\pi} \Theta_{c}} \zeta_R \zeta_L
\nonumber \\ 
\left[\cos\left(\sqrt{2\pi}\bar \Theta_1\right),
-\sin\left(\sqrt{2\pi}\bar \Theta_1\right),
\sin\left(\sqrt{2\pi}\bar \Phi_1\right)\right].
\label{convordmod1}
\eea
All these conventional order parameters have thus
the same scaling dimension ($2$) and the corresponding pair correlation
functions decay as $x^{-4}$ i.e. 
much faster than in the one dimensional metals.
One should notice that the $Z_2$ pseudocharge excitations 
contribute in (\ref{convordmod1}) through the
density energy operator $\ri \zeta_R \zeta_L$. In a critical Ising
model, there are other primary fields (order and disorder parameters)
that have a smaller scaling dimension ($1/8$) which are highly {\sl nonlocal}
in terms of the original Majorana fermion\cite{book}. 
This leads us to investigate 
the possibility that the ground state might be characterized
by composite order parameters: 
staggered odd-frequency singlet pairing (c-SP) and 
staggered composite CDW (c-CDW).
At the IR fixed point, we find the 
following correspondence for these order parameters:
\bea
{\cal O}_{c-SP} &=& {\vec {\cal O}}_{TS}\cdot \left({\bf n}_1 + 
{\bf n}_2\right) 
\sim \re^{-\ri \sqrt{2\pi} \Theta_{c}} \mu_0 \mu_4
\nonumber \\
{\cal O}_{c-CDW} &=& {\vec {\cal O}}_{SDW}\cdot \left({\bf n}_1 +
{\bf n}_2\right)                                              
\sim
\re^{\ri \sqrt{2\pi} \Phi_{c}} \mu_0 \mu_4 
\label{unconvordmod1}
\eea                                                            
where $\mu_0,\mu_4$ (respectively $\sigma_0,\sigma_4$) 
are the Ising disorder (respectively order) parameters
associated with the Majorana fermions $\xi^0,\zeta$.
>From Eq. (\ref{unconvordmod1}), we deduce the leading
asymptotics of the correlation function corresponding to
the composite order parameters:
\bea
\langle {\cal O}_{composite}\left(x,\tau\right)
{\cal O}_{composite}^{\dagger}\left(0,0\right) \rangle \sim
\nonumber \\
\frac{1}{\left(v_F^2 \tau^2 + x^2\right)^{1/2}
\left(v_1^2 \tau^2 + x^2\right)^{1/8}
\left(u_2^2 \tau^2 + x^2\right)^{1/8}}
\label{corcomp}
\eea
so that the composite order parameters induce the 
dominant instabilities and have enhanced long-range
coherence.
These fluctuations can be made even more enhanced 
upon
switching on a very small ferromagnetic ($g_{12} < 0$)
exchange interaction between the spin chains: ${\cal O}_{12}
= g_{12} {\bf S}_1 \cdot {\bf S}_2 $. 
Indeed, at the IR fixed point, this perturbation mostly
affects the pseudocharge sector leaving intact the magnetic
one: ${\cal O}_{12} \simeq g_{12} \ri \xi^0_R \xi^0_L$. 
The Majorana fermion $\xi^0$ acquires a positive mass
i.e. its associated Ising model is in the disorder phase: $\langle \mu_0
\rangle \ne 0$ and the $Z_2$ symmetry with respect to the 
interchange of the two chains is now broken.
All other degrees of freedom remain critical so that 
the pair correlation of the 
conventional order parameters (\ref{convordmod1}) still
decay as $x^{-4}$ whereas from Eq. (\ref{unconvordmod1})
one immediately observes that the composite ones decay 
now as $x^{-5/4}$ instead of $x^{-3/2}$ at $g_{12}=0$.

{\sl The two-channel KHL case.}  
The determination of the ground-state physical properties 
of the two-channel KHL proceeds in the same way as in 
the previous model. We first use 
the Abelian bosonisation of
the right-left moving fermions:
$(R,L)_{a\sigma} = (\kappa_{a\sigma}/{\sqrt{2\pi a_0}})
\; \re^{\ri\tau_{\sigma}\pi/4}
\re^{\pm \ri \sqrt{4\pi} \varphi_{a\sigma R,L}}$
where
$\varphi_{a\sigma R,L} = (\Phi_{acR,L}
+ \tau_{\sigma} \varphi_{aR,L})/\sqrt{2}$, and
$\kappa_{a\sigma}$ are some Klein-factors\cite{comment}.    
The Green's functions of the chiral fermions 
have a similar structure than in Eq. (\ref{greenmod1})
apart from they fall now with the distance with a power $5/4$.
The leading asymptotics of the 
correlation function of the localized spins
have been computed in Ref.\cite{azarial} and in particular the
staggered part decays as $x^{-3}$.
Using the Toulouse basis (\ref{cano}) and introducing
the charge and relative charge bosonic fields 
$\Phi_{\pm cR,L} = (\Phi_{1cR,L} \pm \Phi_{2cR,L})/\sqrt{2}$, 
the order parameters for the electronic degrees of freedom 
can also be expressed in terms of the critical fields at the IR fixed point.
We find the 
following correspondence 
for the conventional 
order parameters:
\bea
{\cal O}_{CDW} = L_{a\sigma}^{\dagger} R_{a\sigma}
\sim 
\re^{\ri \sqrt{\pi} \Phi_{+c}}
\cos\left(\sqrt{2\pi}\bar \Phi_1\right) 
{\cal I}_{\Phi}
\nonumber \\
{\vec {\cal O}}_{SDW} =
 L_{a\alpha}^{\dagger} {\vec \sigma}_{\alpha \beta}R_{a\beta}
\sim
\re^{\ri \sqrt{\pi} \Phi_{+c}} \nonumber \\
\left[\cos\left(\sqrt{2\pi}\bar \Theta_1\right)
{\cal I}_{\Phi}^{\dagger},
\sin\left(\sqrt{2\pi}\bar \Theta_1\right) {\cal I}_{\Phi}^{\dagger},
-\sin\left(\sqrt{2\pi}\bar \Phi_1\right)
{\cal I}_{\Phi}\right]
\nonumber \\
{\cal O}_{SS} = -\ri L_{a\alpha}
\left(\sigma^y\right)_{\alpha \beta}
R_{a\beta} \sim                            
\re^{-\ri \sqrt{\pi} \Theta_{+c}}
\cos\left(\sqrt{2\pi}\bar \Phi_1\right) {\cal I}_{\Theta}^{\dagger}
\nonumber \\
{\vec {\cal O}}_{TS} = -\ri L_{a\alpha}
\left({\vec \sigma} \sigma^y\right)_{\alpha \beta}
R_{a\beta} \sim
\re^{-\ri \sqrt{\pi} \Theta_{+c}}
\nonumber \\
\left[-\cos\left(\sqrt{2\pi}\bar \Theta_1\right)
{\cal I}_{\Theta},
\sin\left(\sqrt{2\pi}\bar \Theta_1\right){\cal I}_{\Theta},
\sin\left(\sqrt{2\pi}\bar \Phi_1\right){\cal I}_{\Theta}^{\dagger}\right]
\nonumber
\eea                                                              
where ${\cal I}_{\Phi} = 
\mu_0 \mu_4 \cos\left(\sqrt{\pi}\Phi_{-c}\right)
+ \ri \sigma_0 \sigma_4 \sin\left(\sqrt{\pi}\Phi_{-c}\right)$
and ${\cal I}_{\Theta}$ is obtained from
${\cal I}_{\Phi}$ by replacing $\Phi_{-c}$
by its dual field $ \Theta_{-c}$.
The scaling
dimension of all these conventional order parameters is $5/4$
and their corresponding pair correlation functions
decay thus as $x^{-5/2}$ i.e. have no enhanced long-range coherence. 
Notice, however, that this decay is much slower than in 
the 1DEG in a special active environment case. 
Similarly,
we find the following description at the IR fixed 
point for the staggered composite order parameters:
\bea
{\cal O}_{c-SP} = {\vec {\cal O}}_{TS}\cdot {\bf n}_0
\sim \re^{-\ri \sqrt{\pi} \Theta_{+c}} {\cal I}_{\Theta}
\nonumber \\
{\cal O}_{c-CDW} = {\vec {\cal O}}_{SDW}\cdot {\bf n}_0
\sim
\re^{\ri \sqrt{\pi} \Phi_{+c}} \nonumber \\
\left(\mu_0 \mu_4
\cos\left(\sqrt{\pi}\Phi_{-c}\right) + 
3 \ri \sigma_0 \sigma_4 \sin\left(\sqrt{\pi}\Phi_{-c}\right)\right)
\label{unconvordmod2}
\eea 
so that the scaling dimension of these operators is $3/4$ 
and the pair correlation function of the composite order 
parameters is still given by Eq. (\ref{corcomp}).
The dominant instabilities of the two-channel KHL 
are thus of a composite nature as in the 1DEG in a special 
active environment. Finally, one should note that
the scaling dimensions of the physical observables 
of the two-channel KHL obtained by the Toulouse point solution
coincide with those of the recent CFT approach\cite{orignac}.
In this respect, we stress that, in contrast with
the latter approach, the  Toulouse point solution 
unables us to take into account the inherent velocities anisotropy in the 
problem. 

In summary,  
we have shown  
that
the two generalizations of the one dimensional KHL 
considered in this letter
exhibit a nontrivial non-Fermi liquid low-temperature 
phase belonging to the class of chirally stabilized fluids 
with strong enhanced staggered composite pairing correlations.

\sloppy
\par


\end{document}